\documentclass[aps,prl,reprint,showpacs,superscriptaddress,longbibliography]{revtex4-1}
\usepackage{mathtools,amssymb,graphicx,units}
\usepackage[usenames,dvipsnames]{color}
\usepackage[plainpages=false,pdfpagelabels,colorlinks=true,linkcolor=red,urlcolor=blue,citecolor=blue,pdftitle={Title},pdfauthor={},pdfdisplaydoctitle=true,pdfduplex=DuplexFlipLongEdge]{hyperref}
\usepackage{subfigure}
\usepackage{grffile}
\usepackage{bm}
\usepackage{mhchem}
\usepackage{color}
\usepackage{hyperref}
\usepackage{bibentry}

\newcommand{\1}{\mbox{\bf 1}}
\begin{document}
\title{High-harmonic generation approaching the quantum critical point of strongly correlated systems}
\author{Can Shao}
\email{shaocan@njust.edu.cn}
\affiliation{Institute of Ultrafast Optical Physics, Department of Applied Physics, Nanjing University of Science and Technology, Nanjing 210094, China}
%\affiliation{Beijing Computational Science Research Center, Beijing 100084, China}

\author{Hantao Lu}
\affiliation{School of Physical Science and Technology $\&$ Key Laboratory for Magnetism and Magnetic Materials of the MoE, Lanzhou University, Lanzhou 730000, China}

\author{Xiao Zhang}
\affiliation{Institute for Theoretical Solid State Physics, Leibniz IFW Dresden, Helmholtzstr. 20, 01069 Dresden, Germany}

\author{Chao Yu}
\affiliation{Institute of Ultrafast Optical Physics, Department of Applied Physics, Nanjing University of Science and Technology, Nanjing 210094, China}

\author{Takami Tohyama}
\affiliation{Department of Applied Physics, Tokyo University of Science, Tokyo 125-8585, Japan}

\author{Ruifeng Lu}
\email{rflu@njust.edu.cn}
\affiliation{Institute of Ultrafast Optical Physics, Department of Applied Physics, Nanjing University of Science and Technology, Nanjing 210094, China}

% \date{\today}

\begin{abstract}
By employing the exact diagonalization method, we investigate the high-harmonic generation (HHG) of the correlated systems under the strong laser irradiation. For the extended Hubbard model on a periodic chain, HHG close to the quantum critical point (QCP) is more significant compared to two neighboring gapped phases (i.e., charge-density-wave and spin-density wave states), especially in low-frequencies. We confirm that the systems in the vicinity of the QCP are supersensitive to the external field and more optical-transition channels via excited states are responsible for HHG. This feature holds the potential of obtaining high-efficiency harmonics by making use of materials approaching to QCP. Based on two-dimensional Haldane model, we further propose that the even- or odd-order components of generated harmonics can be promisingly regarded as spectral signals to distinguish the topologically ordered phases from locally ordered ones. Our findings in this work pave the way to achieve ultrafast light source from HHG in strongly correlated materials and to study quantum phase transition by nonlinear optics in strong laser fields.
\end{abstract}

\maketitle

\paragraph{Introduction.---}\label{sec1}

Quantum phase transitions (QPTs) are of extensive interest in condensed matter physics~\cite{Sachdev2011,Vojta2003} because they happen at zero temperature where the thermal fluctuations vanish and the uncertainty effects in quantum physics are manifested.
In a many-body system, phase transition accompanied with the onset of a local order parameter occurs as a result of competing interactions.
Experimental detections of QPTs are straightforward, which include conductivity, susceptibility or total magnetization in some spin systems~\cite{Osterloh2002,Peng2005}.
However, not all order parameters can be measured by such macroscopic measurements and they are not suitable for closer investigations of the quantum critical point (QCP)~\cite{Zhang2008}.
Instead, the dynamical response functions such as the frequency-dependent optical conductivity provide an important route to investigate the quantum criticality~\cite{Lucas2017,Cha1991,Damle1997,Jurij2005,William2012,Myers2011,Witczak-Krempa2012,Witczak-Krempa2014,Chen2014,Gazit2013,Katz2014,Gazit2014,Witczak-Krempa2016}.
Due to the destruction of quasiparticles and the corresponding abundance of incoherent excitations in the vicinity of the QCP, the systems are expected to be much more sensitive to external perturbations than in the center of a phase\cite{Quan2006,Yuan2007}, especially on short time scales.
Thus, one can expect that nonequilibrium and nonlinear behaviors are relatively active in such systems, which may play a role as promising tools to detect QPT and QCP.

Strong-field-driven dynamics and high-harmonic generation (HHG) are perhaps the most representative examples of nonlinear and nonperturbative optical processes~\cite{Brabec2000,Corkum2007}, which are widely expected to generate the attosecond light sources and provide new ultrafast imaging methods~\cite{Cavalieri2007,Krausz2009}.
HHG has been initially studied in atomic and molecular gas systems, in which a characteristic plateau with a cut-off energy is well explained by the three-step model~\cite{Agostini_2004,Krausz2009,Gallmann2012,Corkum1993,Lewenstein1994,Ishikawa2015}.
Subsequently, HHG observed in ZnO crystal is interpreted by intraband Bloch oscillations~\cite{Ghimire2011}, and
the extended three-step model is proposed~\cite{Vampa2014} and recognized in the community of solid HHG (see~\cite{Kruchinin2018,Huttner2017,Ghimire2019,Chao2019} and references therein).
Recently, the studies have also touched on the mechanisms of HHG and the harmonic plateaus in strongly correlated systems~\cite{Murakami2018,Imai2020,Sandra2020,Murakami2021}. It is theoretically proposed that the high-harmonic spectroscopy can be used to time-resolve non-equilibrium many-body dynamics, such as optically driven phase transition~\cite{Silva2018,Zhu2021}. Experimental observation of photoinduced insulator-to-metal phase transition by time-resolved HHG has also been reported in correlated material vanadium dioxide (VO$_2$)~\cite{Bionta2021}.

In this Letter, different from previous explorations on mechanisms or using intensity characteristics to study non-equilibrium dynamics in strongly correlated systems, we shed new light on searching candidate materials for high-efficiency HHG and detecting topological phase transition (in or out of equilibrium) on the basis of even- or odd-order harmonic signals. We first study the ultrafast dynamics of the half-filled extended Hubbard model on the one-dimensional (1D) chain, based on the exact diagonalization (ED) method. Due to the sensibility of system close to QCP that separates two gapped phases, i.e., the spin-density wave (SDW) and charge-density-wave (CDW) states, an enhancement of the HHG intensity can be observed. Meanwhile, the HHG spectroscopy has a good correspondence with the optical conductivity in equilibrium. This might provide a new insight to explain the HHG plateau and cut-off energy in correlated systems. In two-dimension (2D), a topological phase transition from Chern insulator (CI) to the CDW phase occurs in the interacting Haldane model. Different from that only odd-order components of HHG appear in CDW phase, both odd and even harmonic orders exist in CI phase. This feature can be utilized to detect topological phase transition in or out of equilibrium.

\paragraph{Models and observables.---}\label{sec_model}
We consider two models to calculate the HHG: the spinful extended Hubbard model and the spinless Haldane model with nearest-neighbour interactions, both at half filling. The former is defined on a periodic chain, which reads
\begin{eqnarray}
\hat H&=&-t_1\sum_{\langle i,j\rangle,\sigma}\left(\hat c^{\dagger}_{i,\sigma} \hat c_{j,\sigma}+\text{H.c.}\right)+U\sum_{i}\left(\hat n_{i,\uparrow}-\frac{1}{2}\right)\nonumber\\
&&\times\left(\hat n_{i,\downarrow}-\frac{1}{2}\right)
+V\sum_{\langle i,j\rangle}\left(\hat n_{i}-1\right)\left(\hat n_{j}-1\right),
\label{H}
\end{eqnarray}
where $\hat c^{\dagger}_{i,\sigma}$ ($\hat c_{i,\sigma}$) creates (annihilates) an electron at site $i$ with spin $\sigma=\uparrow,\downarrow$, and $\hat n_i= \hat n_{i,\uparrow}+ \hat n_{i,\downarrow}$ is the number operator of electrons; $t_1$ is the hopping constant; $U$ and $V$ are the strengths of the on-site and nearest-neighbor (NN) Coulomb-interactions, respectively. The lattice size is set to be $L = 10$.

On the honeycomb lattice, we study the half-filled spinless Haldane model with repulsive NN interactions:
\begin{eqnarray}
\hat H=&-&t_1\sum_{\langle i,j\rangle}(\hat c^{\dagger}_{i} \hat c^{\phantom{}}_{j}+\text{H.c.})
-t_2\sum_{\langle\langle i,j\rangle\rangle}(e^{{\rm i}\phi_{ij}}\hat c^{\dagger}_{i} \hat c^{\phantom{}}_{j}+\text{H.c.}) \nonumber \\
&+&V\sum_{\langle i,j\rangle}\hat n_{i}\hat n_{j}.
\label{eq:H}
\end{eqnarray}
$t_1$ and $t_2$ are the NN and next-nearest-neighbor (NNN) hopping constants,respectively. Same as before, $V$ represents the NN interaction strength. A phase $\phi_{ij}=\frac{\pi}{2}$ ($-\frac{\pi}{2}$) in the anticlockwise (clockwise) loops is added to the second hopping term, which breaks the time-reversal symmetry and turns the system to be topologically non-trivial.

%\shaox{To explain the intensity of the high-harmonic spectroscopy in correlated systems, }
We calculate the real part of the optical conductivity in equilibrium, which is given by the Kubo formula:
\begin{eqnarray}
\text{Re}\ \sigma(\omega)=\frac{\pi}{L}\sum_{m\neq0}|\langle\psi_m|\hat J|\psi_0\rangle|^2 \delta(\omega+E_m-E_0)
\label{eq:2}
\end{eqnarray}
where $|\psi_0\rangle$ and $|\psi_m\rangle$ are the ground state and $m$-th eigenstate, respectively. Eq. (\ref{eq:2}) only gives the the optical conductivity with finite frequency because $m\neq0$. The delta function is broaden by using a Lorentzian shape with a broadening factor $\eta=0.1$. The current operator on the 1D chain reads
\begin{eqnarray}
\hat J=-it_1\sum_{\langle i,j\rangle,\sigma}[\hat c_{i,\sigma}^{\dagger}\hat c_{j,\sigma}-\text{H.c.}],
\label{eq:3}
\end{eqnarray}
while on the 2D honeycomb lattice we have
\begin{eqnarray}
\hat J_x=&-&it_1\sum_{\langle i,j\rangle,\sigma}\textbf{R}_{ij}\cdot\mathbf{e}_{x}\ [\hat c_{i,\sigma}^{\dagger}\hat c_{j,\sigma}-\text{H.c.}] \nonumber \\
&-&it_2\sum_{\langle \langle i,j\rangle\rangle,\sigma}\textbf{R}_{ij}\cdot\mathbf{e}_{x}\ [e^{{\rm i}\phi_{ij}}c_{i,\sigma}^{\dagger}c_{j,\sigma}-\text{H.c.}],
\label{eq:4}
\end{eqnarray}
where $\textbf{R}_{ij}=\textbf{R}_{j}-\textbf{R}_{i}$ and the $x$ direction is defined to be along the nearest-neighbour sites.

Out of equilibrium, we adopt the time-dependent Lanczos technique in ED to evolve the many-body wave function, see Supplemental Material~\cite{Supple}.
The external electric field during photoirradiation can be included into the Hamiltonian via the Peierls substitution in the hopping terms:
\begin{eqnarray}
\hat c^{\dagger}_{i,\sigma}\hat c_{j,\sigma}+\text{H.c.}\rightarrow
e^{\mathrm{i}\textbf{A}(t)\cdot(\textbf{R}_j-\textbf{R}_i)}\hat c^{\dagger}_{i,\sigma}\hat c_{j,\sigma}+\text{H.c.},
\label{eq:Peierls}
\end{eqnarray}
where $\textbf{A}(t)=(A_x(t),A_y(t))$ is the vector potential and
\begin{eqnarray}
A_{x}(t)=\left\{
\begin{aligned}
&A_{0,{x}}e^{-t^2/2t_d^2}\cos\left(\omega_0 t\right),\ t<0\\
&A_{0,{x}}\cos\left(\omega_0 t\right),\ t\geq0
\end{aligned}
\right.
\end{eqnarray}
\begin{eqnarray}
A_{y}(t)=\left\{
\begin{aligned}
&A_{0,{y}}e^{-t^2/2t_d^2}\sin\left(\omega_0 t\right),\ t<0\\
&A_{0,{y}}\sin\left(\omega_0 t\right),\ t\geq0.
\end{aligned}
\right.
\end{eqnarray}
The parameter $t_d$ controls the width of the Gaussian-like envelope with $t<0$ and $\omega_0$ is the fundamental frequency of incident light. In the 2D case, we set $A_{0,x}=A_{0,y}$ to simulate the circularly polarized laser, while in the case of 1D chain, we set $A_{0,y}=0$ ($A_{0}=A_{0,x}$) to simulate the linearly polarized one. The time-dependent current density is defined as $\langle j\rangle_t=\langle\psi(t)|\hat J|\psi(t)\rangle/L$ or $\langle j_x\rangle_t=\langle\psi(t)|\hat J_x|\psi(t)\rangle/A_s$ accordingly, where L is the number of lattice of the chain and $A_s$ is the total area of the honeycomb lattice. We have to stress that the Peierls substitution in Eq.~(\ref{eq:Peierls}) must be also added to the current operator in Eq.~(\ref{eq:3}) and (\ref{eq:4}) out of equilibrium. The HHG spectrum $|\langle j\rangle_{\omega}|^{2}$ is obtained as the modulus square of the Fourier transform of the time-dependent current density $\langle j\rangle_t$ or $\langle j_x\rangle_t$.

In this letter, we use a set of the natural units for the description of electromagnetic field and related quantities, taking the reduced Planck constant $\hbar$, the elementary charge $e$, the light velocity $c$ and the lattice constant $a_0$ to be $1$. Meanwhile, $t_1$ and ${t_1}^{-1}$ are the units of energy and time, respectively. Taking the relevant material ET-F$_2$TCNQ as an example, we demonstrate the realistic units of time and energy  as well as the feasibility of laser parameters in practical experiments in Supplemental Material~\cite{Supple}.

\paragraph{Results.---}\label{sec_Results}

\begin{figure}[t]
\centering
\includegraphics[width=0.45\textwidth]{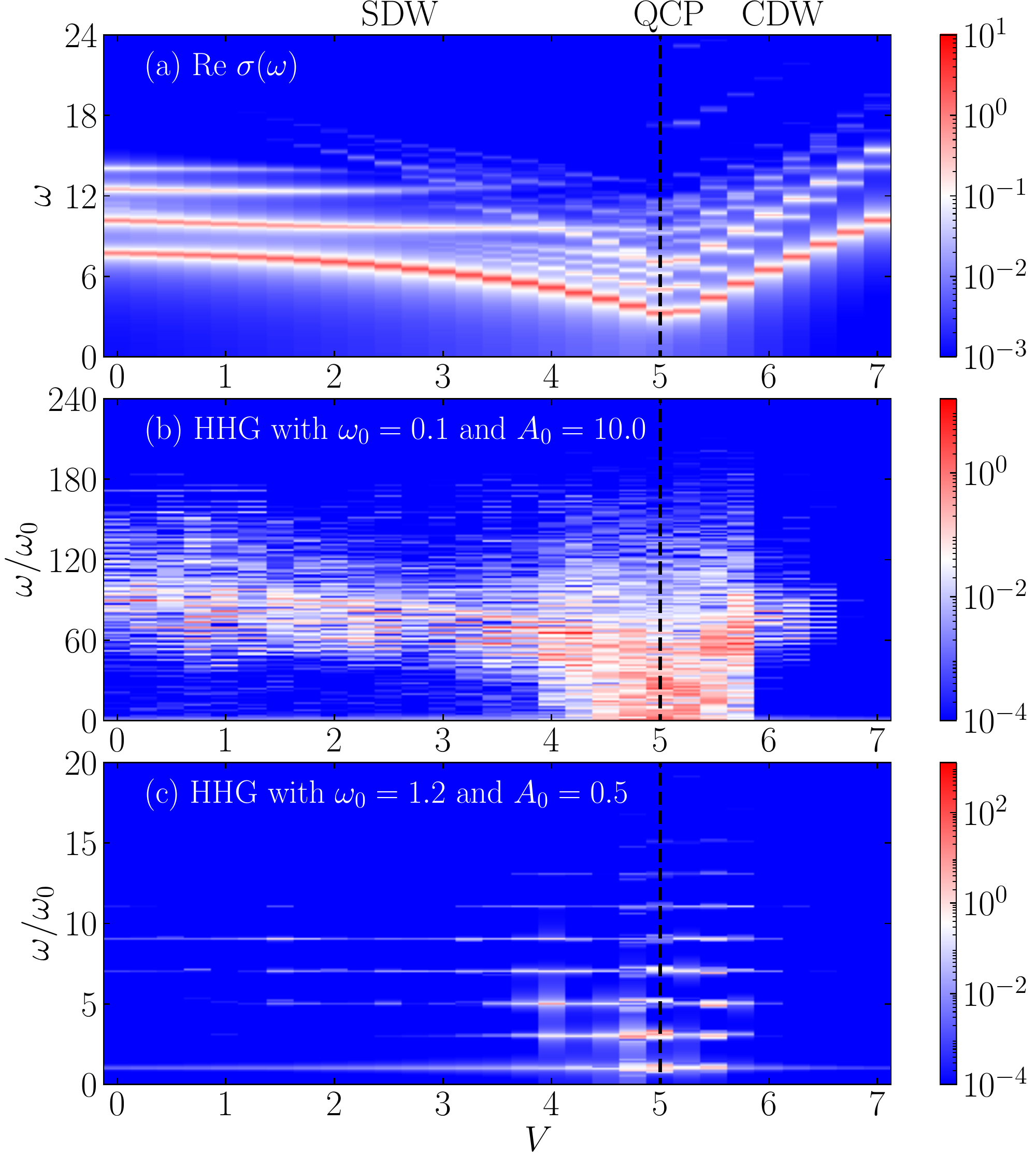}
\caption{(a) Contour plots of the optical conductivity Re $\sigma(\omega)$ as a function of $\omega$ and the NN interactions $V$. Contour plots of HHG spectrum $|\langle j\rangle_{\omega}|^{2}$ as a function of $\omega/\omega_0$ and $V$, with $\omega_0=0.1$ and $A_0=10.0$ in (b) as well as $\omega_0=1.2$ and $A_0=0.5$ in (c). Other parameters of the Hamiltonian (\ref{H}) and the external laser are set to be $U=10.0$ and $t_d=50.0$.}
\label{fig_1}
\end{figure}

We set $U=10.0$ for the $1$D extended Hubbard model and the phase transition between SDW and CDW locates at $V\simeq U/2=5.0$~\cite{Ejima2007}.
Figure \ref{fig_1} (a) shows the optical conductivity Re $\sigma(\omega)$ in equilibrium with changing the NN interaction $V$, where we can observe a minimum optical gap at $V=5.0$. Such features have also been studied in Ref. \cite{Shao2021}, together with a minimum of the single-particle gap.

The HHG spectrum $|\langle j\rangle_{\omega}|^{2}$ as a function of $\omega/ \omega_0$ and the interaction $V$ are plotted in Figs.~\ref{fig_1} (b) and (c), with $\omega_0=0.1$, $A_0=10.0$ and $\omega_0=1.2$, $A_0=0.5$, respectively. Interestingly, there is an obvious enhancement of HHG spectrum approaching to the critical point $V=5.0$. This can serve as an optical tool to detect the QCP between two insulating phases, which can not be directly measured by the traditional electrical methods. In SDW and CDW phases, the harmonic orders with high intensity of HHG have a very good correspondence with the optical conductivity through multiplied by the fundamental frequency $\omega_0$.
%Difference is that in CDW regime, the HHG spectrum loses its weight gradually with the increasing of $V$.
From the definition of Eq. (\ref{eq:2}), we know that the spectra of optical conductivity are associated with the corresponding excited states that can be connected to the ground state by the current operator. These excited states are called the optically allowed states. The HHG is a kind of nonlinear process with absorbing $m$ multiples of photons and generating laser with frequency $m$ multiples of the incident light. Thus, the integer $m$ strongly depends on the energy difference between the ground state and the optically allowed states, which explains the similarity between the optical conductivity and HHG spectrum.
This feature may provide a new way to predict the HHG plateau and cut-off energy in correlated materials. Deep in SDW and CDW phases, the generated harmonics (such as the 3rd and 5th orders in Fig.~\ref{fig_1} (c)) are suppressed, while they are enhanced when the system becomes closer to QCP, which can be well explained by the flatness of the band structures in SDW and CDW phases (See details in Supplemental Material \cite{Supple}).
We propose that such an intriguing phenomenon could be utilized to generate HHG with higher strength. In addition, the fact that intensity of HHG in CDW phase with larger $V$ becomes more and more weak can be attributed to the rapid increment of the optical gap, i.e., the energy difference between the ground state and the lowest optical allowed excited state.

\begin{figure}[t]
\centering
\includegraphics[width=0.45\textwidth]{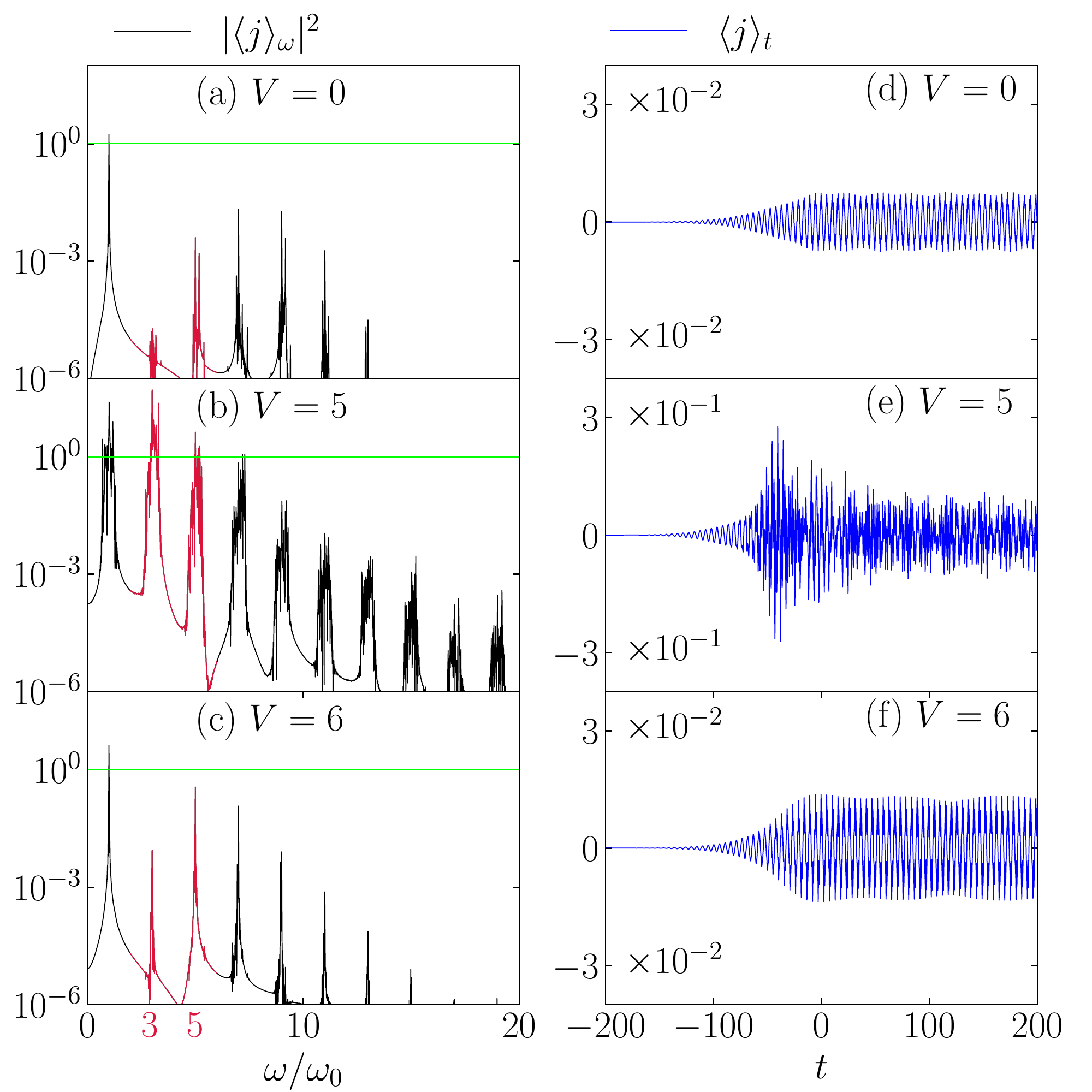}
\caption{The HHG spectrum $|\langle j\rangle_{\omega}|^{2}$ as a function of $\omega/\omega_0$ with $V=0.0$ (a), $V=5.0$ (b) and $V=6.0$ (c). The 3rd and 5th harmonics are plotted in red to emphasize the difference in intensity. Time profiles of $A(t)$ (red lines) and $\langle j\rangle_t$ (blue lines) with $V=0.0$ (d), $V=5.0$ (e) and $V=6.0$ (f). Other parameters of the Hamiltonian (\ref{H}) and the external laser are set to be $U=10.0$, $\omega_0=1.2$, $A_0=0.5$ and $t_d=50.0$.}
\label{fig_2}
\end{figure}

Now we start to discuss the details of the HHG spectrum and the ultrafast dynamics of 1D extended Hubbard model. We choose $V=0$ and $V=6$ in SDW and CDW phase, respectively, and $V=5$ very close to the critical point to plot $|\langle j\rangle_{\omega}|^{2}$ as a function of $\omega/\omega_0$, see Figs.~\ref{fig_2} (a), (b) and (c). Parameters of the incident laser are identical to those in Fig.~\ref{fig_1} (c). In order to obtain $|\langle j\rangle_{\omega}|^{2}$, we do the Fourier transform of $\langle j\rangle_t$ from $t=-300$ to $t=400$. We observe that all the harmonic components of HHG spectrum locates at $\omega/\omega_0=2n+1$ with $n>0$ and the lower-order ones ($3$rd and $5$th harmonic order) with $V=5.0$ are much stronger than those in the other two cases, i.e., $V=0.0$ and $V=6.0$. However, one can not clearly observe the sharp peaks of HHG with $V=5.0$, which we speculate is due to the rapid heating process near the critical point (the electronic thermalization of this model has been discussed in Ref.~\cite{Shao2019}). To examine this idea, we plot the time evolution of the current density $\langle j\rangle_t$ in Figs.~\ref{fig_2} (d), (e) and (f). We can find a quick and intense current response in the case of $V=5.0$ with the order of magnitude being $10^{-1}$ when the light starts to pump in. As the light shinning steadily ($t>0$), an obvious suppression of the current response takes place and irregular current-density oscillations appear in Fig.~\ref{fig_2} (e), which are responsible for the indistinguishable peaks in its HHG spectrum. Based on the fact that timescale of the electron-phonon scattering is much larger than that of electron-electron interaction, we thus ignore the energy dissipation from electrons to the phonon bath. The photoinduced energy accumulation leads to the heating of our electronic system and the irregular current oscillations (see more discussions in Supplemental Material \cite{Supple}). The reference \cite{Wang2021} also reported that the peaks of their odd-order harmonics get cleaner by introducing the imaginary potential to phenomenologically depict the dephasing process in the solid HHG. %\shaox{While in our case, there is no such channels to dissipate the injected energy, which cause the consequent heating process. In addition, current responses of $\langle j\rangle_t$ inside SDW and CDW phase seem to be regular with one order of magnitude smaller than the case of $V=5$.}

\begin{figure}[t]
\centering
\includegraphics[width=0.45\textwidth]{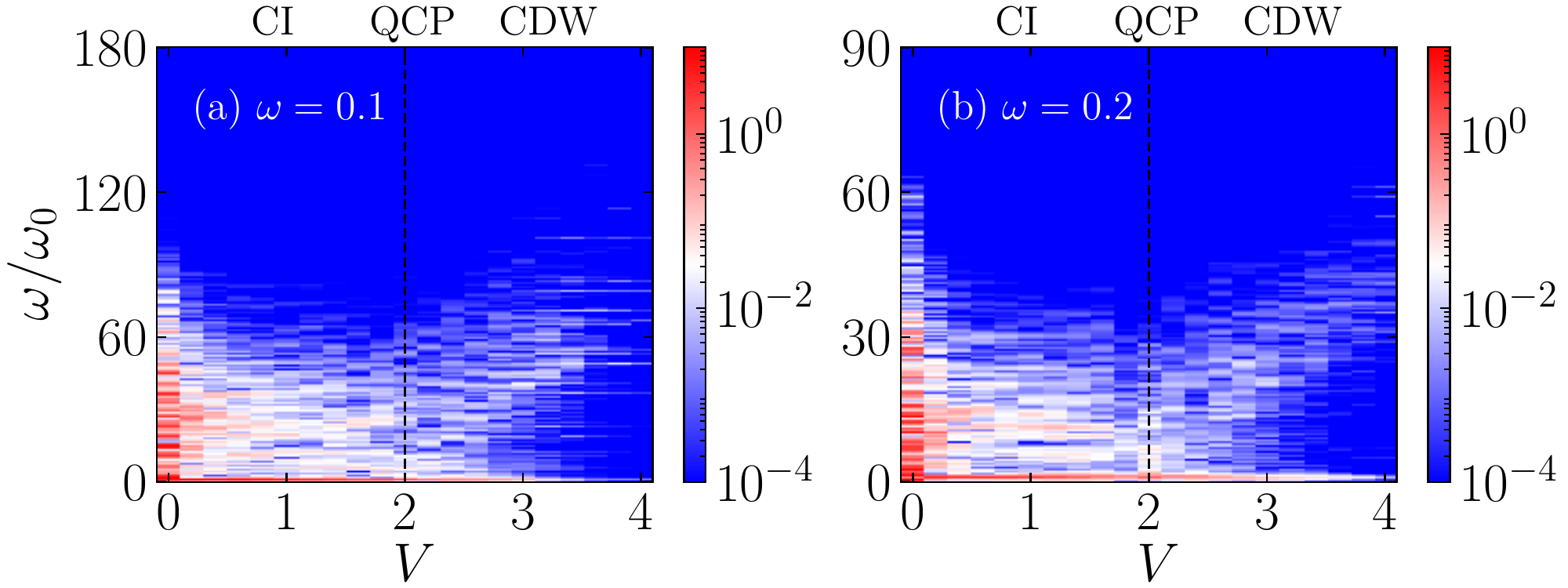}
\caption{Contour plots of HHG spectrum $|\langle j\rangle_{\omega}|^{2}$ as a function of $\omega/\omega_0$ and $V$, with $\omega_0=0.1$ in (a) and $\omega_0=0.2$ in (b). Other parameters of the Hamiltonian (\ref{eq:H}) and the external laser are set to be $t_2=0.2$, $A_{0,x}=10.0$, $A_{0,y}=10.0$ and $t_d=50.0$.}
\label{fig_4}
\end{figure}

\begin{figure}[t]
\centering
\includegraphics[width=0.45\textwidth]{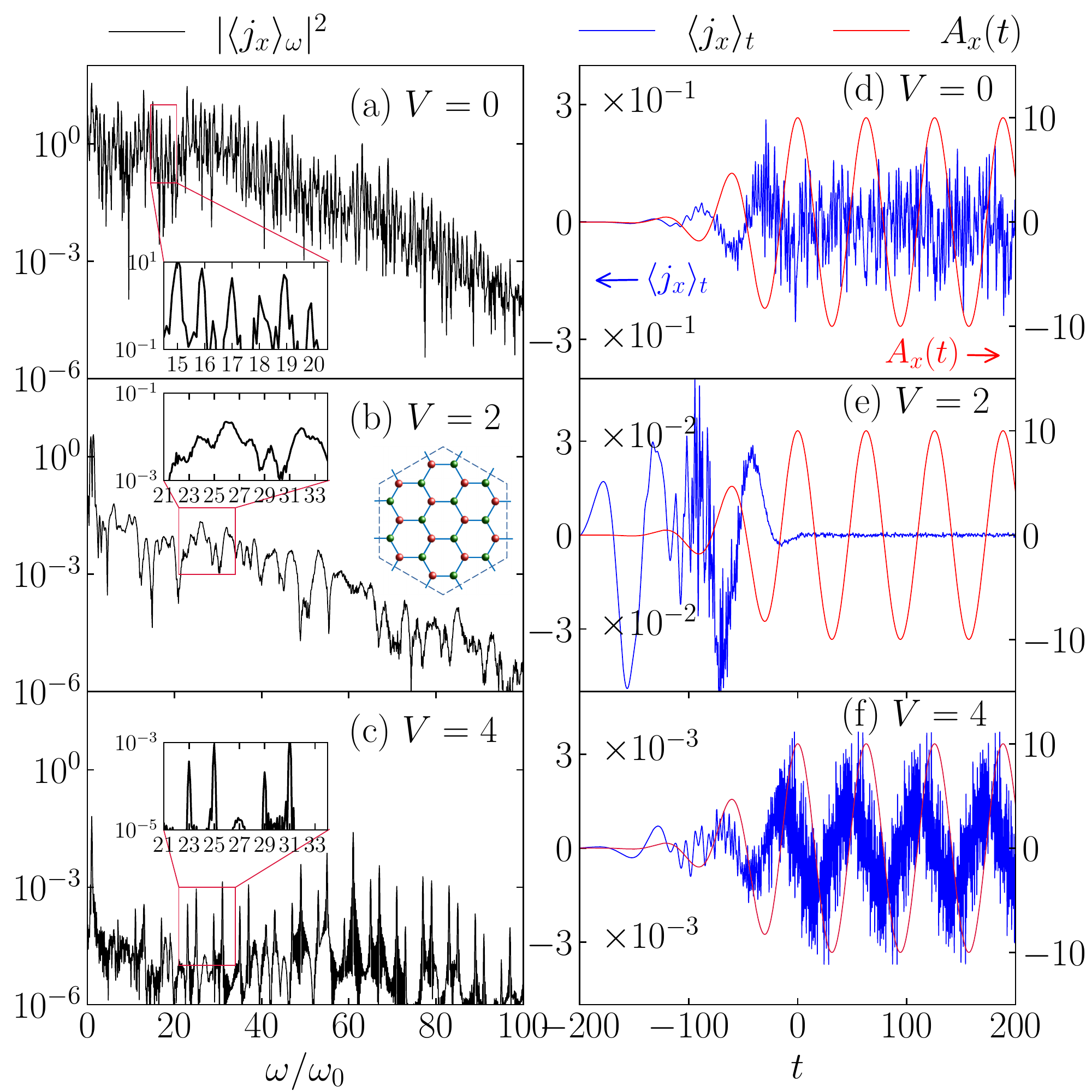}
\caption{The HHG spectrum $|\langle j\rangle_{\omega}|^{2}$ as a function of $\omega/\omega_0$ with $V=0.0$ (a), $V=2.0$ (b) and $V=4.0$ (c). Time profiles of $A_x(t)$ (red lines) and $\langle j_x\rangle_t$ (blue lines) with $V=0.0$ (d), $V=2.0$ (e) and $V=4.0$ (f). Other parameters of the Hamiltonian (\ref{eq:H}) and the external laser are set to be $t_2=0.2$, $\omega_0=0.1$, $A_0=10.0$ and $t_d=50.0$.}
\label{fig_5}
\end{figure}

For the interacting spinless Haldane model (\ref{eq:H}), the topological phase transition from a Chern insulator (CI) towards a trivial CDW insulator with growing interactions has been studied by Varney et al \cite{Varney2010,Varney2011}. Here we adopt the $24$A lattice with periodic boundary condition shown in the inset of Fig.~\ref{fig_5} (b), which can largely reduce the finite-size effect because of its good symmetry \cite{Varney2011}. We set $t_2=0.2$ and the QCP locates at $V\approx2.0$. Contour plots of the HHG spectrum $|\langle j\rangle_{\omega}|^{2}$ as a function of $\omega/\omega_0$ and $V$ are shown in Figs.~\ref{fig_4} (a) and (b), with the incident laser frequency $\omega_0=0.1$ and $\omega_0=0.2$, respectively. Other parameters of the external circularly polarized laser are set to be $A_{0,x}=10.0$, $A_{0,y}=10.0$ and $t_d=50.0$. Instead of enhancement of HHG close to QCP, we observe a gradually decreasing of the HHG intensity with $V$ increasing. So we speculate this is due to the gapless CI phase and there are already enough low-energy excited states to contribute the harmonic generation in the topological phase. This supports that the topological edge states inside the bulk gap might favor a stronger HHG.

%\begin{figure}[b]
%\centering
%\includegraphics[width=0.3\textwidth]{Figure_3.pdf}
%\caption{(a) Contour plots of the optical conductivity Re $\sigma(\omega)$ as a function of $\omega$ and the NN interactions $V$. Contour plots of HHG spectrum $|\langle j\rangle_{\omega}|^{2}$ as a function of $\omega/\omega_0$ and $V$, with $\omega_0=0.1$ and $A_0=10.0$ in (b) as well as $\omega_0=0.2$ and $A_0=5.0$ in (c). Other parameters of the two-leg extended Hubbard model and the external laser are set to be $U=10.0$ and $t_d=50.0$.}
%\label{fig_3}
%\end{figure}

To see more details of the electron dynamics, we show the time-evolution of $\langle j_x\rangle_t$ for different $V$ in the right panel of Fig.~\ref{fig_5}. With $V$ increasing from $0$ to $4$, amplitudes of the current-density responses decrease from the order of $10^{-1}$ to $10^{-3}$, as shown in Figs.~\ref{fig_5} (d), (e) and (f). This results in a weaker HHG intensity for larger $V$, as seen in Figs.~\ref{fig_5} (a), (b) and (c). Similar to the 1D case, there is a apparent suppression of current response occurring soon after applying the light to system with $V=2.0$ and the heating process in the 2D case comes more rapidly and completely.
%We think there are more lower-energy excited states serving as the channels to thermalize the system.
By inspecting Figs.~\ref{fig_5} (a) and (c) as well as their subplots carefully, we find that there are both odd- and even-order components of HHG when $V=0$ in the CI phase, while most harmonic order in CDW side are odd numbers with a suppression of the peaks for the number $3\times(2n+1)$. Such $3\times(2n+1)$ peaks could be observed by adopting another shape of $24$-site lattice (see Supplemental Material~\cite{Supple}), but the even-order number peaks can not be revisited by changing shape or lattice size. So we propose that the odd- or even-order components of HHG spectrum can be utilized to distinguish topologically and locally ordered states.

\paragraph{Summary and discussion.---}\label{conclusion}
Quantum phase transition and its critical behavior are playing an important role in the field of condensed matter physics. By studying the extended Hubbard model on the periodic chain, we found that the optical-allowed excited states, which can be measured by optical conductivity in equilibrium, contribute the formation of HHG spectrum. When the system is close to the critical point which separates two gapped phases, more intense HHG especially in low frequencies is observed because there are more optical allowed excited states. Such phenomenon can be reproduced in the same model on a two-leg ladder, see Supplemental Material for more details~\cite{Supple}. For the interacting Haldane model on the honeycomb lattice, enhancement of HHG close to the topological phase transition point is not observed because the original CI phase is gapless. However, the odd- or even-order components of HHG spectrum provide another way to detect the QPT.

The issue remains open about whether the enhancement of HHG intensity can be accessed in ultrafast experiments for some materials. The candidates include the quasi-1D organic Mott insulators of the TCNQ family~\cite{Uemura2008}, in particular ET-F$_2$TCNQ which is widely studied because of the existence of both on-site and NN Coulomb repulsions ($t_1\sim0.1$ eV, $U\sim 1$ eV; refs.~\cite{Wall2011,Hasegawa2000}). In addition, the search can be extended to ladder or 2D materials at half-filling with strong electron correlations, such as Sr$_{14-x}$Ca$_x$Cu$_{24}$O$_{41}$~\cite{Fukaya2015,Osafune1997}.
~\\

\begin{acknowledgments}

C.S. acknowledges support from the National Natural Science Foundation of China (NSFC; Grant No.~12104229). H.L. acknowledges support from NSFC (Grants No.~11874187 and No.~12174168). T.T. is partly supported by CREST, the Japan Science and Technology Agency (Grant No.~JPMJCR1661) from Ministry of Education, Culture, Sports, Science, and Technology, Japan. R.F. acknowledges supports from NSFC of China (Grants No.~11974185) and the Natural Science Foundation of Jiangsu Province (Grant No.~BK20170032).
We would like to thank the referees for inspiring us to connect the laser parameters to experimental condition in reality and deepening our understanding of HHG in correlated systems.

\end{acknowledgments}

%\bibliographystyle{apsrev4-1}
%\bibliography{lt}
%merlin.mbs apsrev4-1.bst 2010-07-25 4.21a (PWD, AO, DPC) hacked
%Control: key (0)
%Control: author (72) initials jnrlst
%Control: editor formatted (1) identically to author
%Control: production of article title (-1) disabled
%Control: page (0) single
%Control: year (1) truncated
%Control: production of eprint (0) enabled
%

\clearpage

\appendix
\section{Supplemental Material (I): Time-dependent Lanczos method}

For the time-dependent Hamiltonian $\hat{H}(t)$ , we apply the time-dependent Lanczos method to evolve the time-dependent wave function $|\psi(t)\rangle$ starting from the initial ground state~\cite{Prelovsek}, via
\begin{equation}
|\psi(t+\delta{t})\rangle\simeq\sum_{l=1}^{M}{e^{-{\mathrm i}\epsilon_l\delta{t}}}|\phi_l\rangle\langle\phi_l|\psi(t)\rangle,
\label{eq:lanczos}
\end{equation}
where $\epsilon_l$ and $|\phi_l\rangle$ are eigenvalues and eigenvectors of $\hat{H}(t)$, respectively, in the Krylov subspace; $M$ is the dimension of the Lanczos basis, and $\delta{t}$ is the time stepping. We select $M=30$ and $\delta{t}=0.02$ to ensure the convergence of numerical evolution within $t\leq700$.
The validity of this method has been checked in numbers of references, such as Ref.~\cite{Shao16}.

\section{Supplemental Material (II): The realistic units and laser parameters for material ET-F$_2$TCNQ}

In the main text, we set $t_1=1$ and $U=10$ for the one-dimensional ($1$D) extended Hubbard model and use the units $a_0=e=\hbar=c=1$, where $a_0$, $e$, $\hbar$ and $c$ are the lattice constant, the elementary charge, the reduced Planck constant and the speed of light, respectively. To provide experimental researchers with more details, we now discuss the realistic units and laser parameters for the example material ET-F$_2$TCNQ, whose hopping constant $t_1\approx0.1$ eV~\cite{Wall2011}, as follows:

1) The energy $\hbar\omega$ is in units of $t_1$. For instance, the optical conductivity Re $\sigma(\omega)$ with $V=5$ in Fig.~1 of the main text has a gap around $\omega\approx3.2$, so the realistic optical gap $\hbar\omega=3.2\times 0.1$ eV $\approx0.32$ eV.

2) The time $t$ is in units of $t_1^{-1}$. So the realistic time unit for this material is $\frac{\hbar}{t_1}=\frac{1.05457266\times10^{-34} \text{J}\cdot \text{s} }{0.1\times 1.6021766208\times 10^{-19} \text{J}}  \approx6.58 \times 10^{-15}$ s.

3) Wavelength $\lambda$ of the incident laser. Let's take $\omega_0=1.2$ as an example, the oscillation period of laser $T=\frac{2\pi}{\omega_0}\cdot\frac{\hbar}{t_1}\approx3.445\times 10^{-14}$s and thus $\lambda=c\cdot T \approx 10 \ $$\mu$m.

4) Continuous laser power $I_0=\frac{1}{2}\ c\ \varepsilon_0E_0^2$, where $E_0$ is the amplitude of electric field. From that $E=\frac{1}{c}\cdot\frac{\partial A}{\partial t}$ and $A=A_0\cdot \text{cos}(\omega_0 t)$, we can obtain $E_0=\frac{\omega_0}{c}\cdot A_0$. Due to the fact that $A_0$ is in units of $\frac{\hbar c}{ea_0}$, the realistic electric-field intensity $E_0=\frac{\hbar\omega_0}{ea_0}\cdot A_0$. Let's set $\omega_0=1.2$ and $A_0=0.5$, one has $\hbar\omega_0=1.2\cdot t_1=0.12$ eV and $\frac{\hbar\omega_0}{e}=0.12$ V. With the lattice spacing $a_0\approx 5.791\times10^{-10}\text{m}$~\cite{HASEGAWA97}, we finally obtain that $E_0=\frac{\hbar\omega_0}{ea_0}\cdot A_0\approx1.0\times10^{8}$ V/m and $I_0=\frac{1}{2}\ c\ \varepsilon_0E_0^2\approx3.64\times10^{10}$ W/cm$^2$.

Compared with the experimental parameters from the relevant references~\cite{Zaks12,Vampa2015,Hohenleutner2015,Langer2016}, we confirm that such a laser with power density $\sim10^{10}$ W/cm$^2$ at the wavelength $\sim10$ $\mu$m can be readily prepared.

\begin{figure}[t]
\centering
\includegraphics[width=0.45\textwidth]{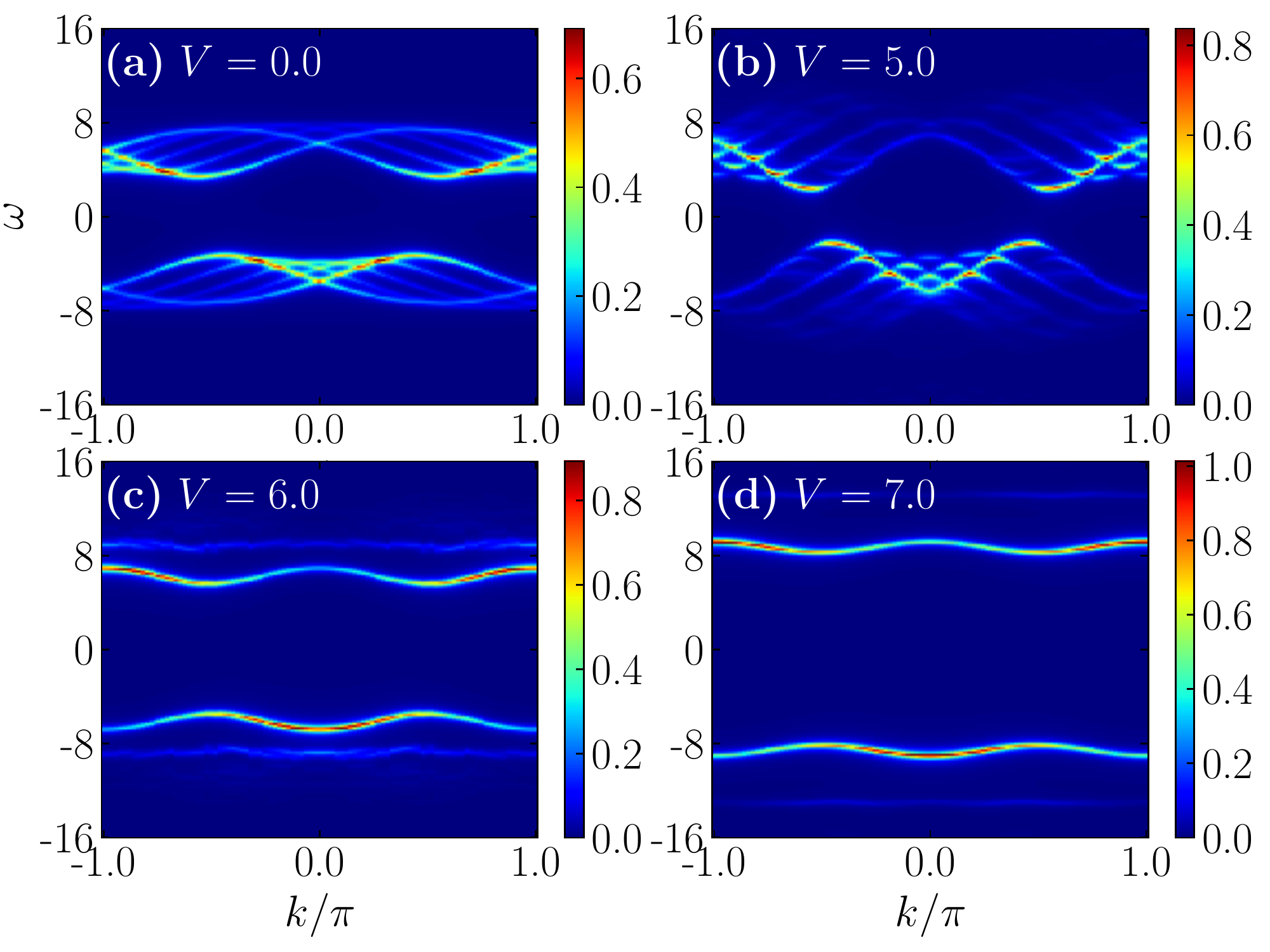}
\caption{The single-particle spectral function $I(k,\omega)$ of the $1$D extended Hubbard model with (a) $V=0.0$, (b) $V=5.0$, (c) $V=6.0$ and (d) $V=7.0$, respectively. We set $U=10.0$ so that the phase transition point is around $V=5.0$.}
\label{fig_band}
\end{figure}

\section{Supplemental Material (III): Band-structure analysis of the one-dimensional extended Hubbard model}\label{thermalization}

The single-particle spectral function $I(k,\omega)$ can be regarded as an effective band structure in interacting systems. Here we produce $I(k,\omega)$ of the $1$D extended Hubbard model with $U=10.0$ for different values of $V$ in Fig.~\ref{fig_band}. Compared with $V=5.0$ in the vicinity of phase transition point, flatter band structures and larger gaps can be observed inside the SDW ($V=0.0$) and CDW ($V=6.0$ and $V=7.0$) phases. According to the three-step-like model of understanding high-harmonic generation (HHG) in solid materials, the lower- and higher-order HHG can be attributed to the intraband Bloch oscillations and interband recombinations of electrons and holes, respectively. The fact that flat band structure hinders intraband oscillations to some extent explains why the lower-order HHG components are largely suppressed in SDW and CDW phases, while they are enhanced when the system is close to the quantum critical point (QCP).

\begin{figure}[t]
\centering
\includegraphics[width=0.45\textwidth]{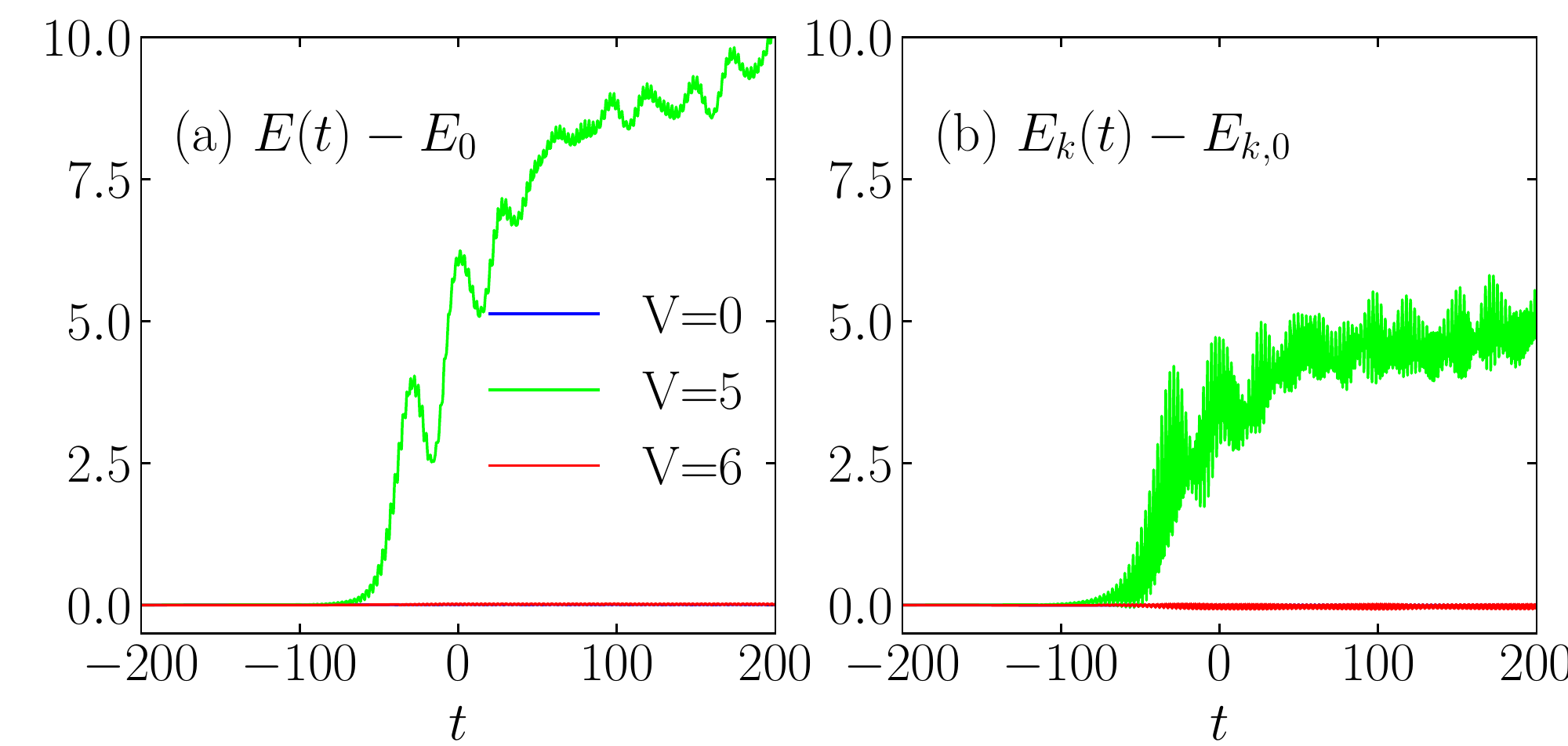}
\caption{The time-dependent total energy (a) and kinetic energy (b) of the $1$D extended Hubbard model with different $V$ under the irradiation. Parameters: $U=10.0$, $\omega_0=1.2$, $A_0=0.5$ and $t_d=50.0$.}
\label{fig1}
\end{figure}

\section{Supplemental Material (IV): Analysis of the injected energy in one-dimensional extended Hubbard model}\label{thermalization}

We examine the ultrafast dynamics of the extended Hubbard model on a periodic chain in the main text and find the current response of system close to the critical point is larger than that in SDW or CDW phase. However, there is an apparent suppression of the current response soon after shinning the light and then the irregular current oscillation dominates. Such an anomalous behavior of current response results in a HHG spectrum with non-clear peaks and the reason is due to the rapid heating of the system. Here we provide the increases of the total energy [$E(t)-E_0$] and the kinetic energy [$E_k(t)-E_{k,0}$] as functions of the time $t$ in Figs.~\ref{fig1} (a) and (b), respectively, where $E(t)=\langle\psi(t)|H(t)|\psi(t)\rangle$ and $E_0$ is the ground-state energy of the Hamiltonian in the absence of the external field. $E_k(t)$ and $E_{k,0}$ are the kinetic part of $E(t)$ and $E_0$, respectively. We can observe that the system with $V=5$ close to the QCP absorbs much more energy (especially the kinetic energy) than in the other two phases. The rapid increase of the kinetic energy happens at $t=-20$, which coincides with the appearance of the irregular current oscillation (see Fig. 2 (e) in the main text). In addition, the kinetic energies in SDW and CDW are observed to oscillate periodically around zero due to the regular driving of the external field. So we propose that close to the critical point, the rapid heating process is responsible for the suppression of current response.

\begin{figure}[t]
\centering
\includegraphics[width=0.45\textwidth]{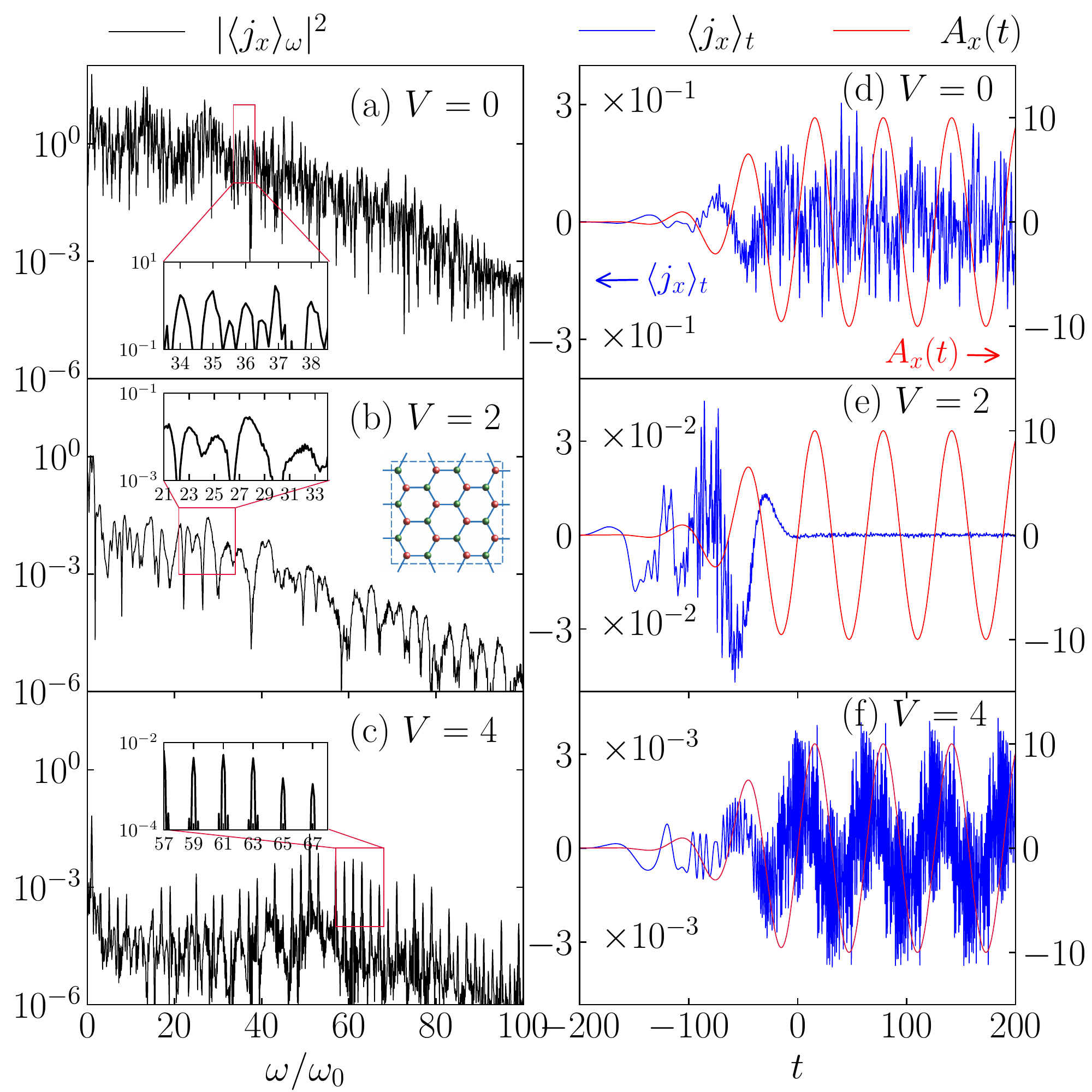}
\caption{The HHG spectrum $|\langle j_x\rangle_{\omega}|^{2}$ as a function of $\omega/\omega_0$ with $V=0.0$ (a), $V=2.0$ (b) and $V=4.0$ (c). Time profiles of $A_x(t)$ (red lines) and $\langle j_x\rangle_t$ (blue lines) with $V=0.0$ (d), $V=2.0$ (e) and $V=4.0$ (f). Other parameters of the interacting Haldane model and the external laser are set to be $t_2=0.2$, $\omega_0=0.1$, $A_0=10.0$ and $t_d=50.0$.}
\label{fig2}
\end{figure}

\section{Supplemental Material (V): HHG spectrum of the interacting Haldane model on the $24$C honeycomb lattice}

In the main text, we show that on the $24$A honeycomb lattice, odd-order HHG with a suppression of the peaks with number $3(2n+1)$ is manifested in CDW phase of the Haldane model with NN interactions. In this section, to check the finite-size effect, a $24$C lattice shown in Fig.~\ref{fig2} (b) is adopted. We do the same calculation with Fig. 4 in the main text. Similarly, we can find that both odd- and even-order harmonic generations can be observed in the Chern insulator with $V=0$, as shown in Fig.~\ref{fig2} (a) and its inset. The rapid heating process also occurs when the system is close to the critical point $V=2$ [see Fig.~\ref{fig2} (e)], which leads to a weak and unclear order of HHG shown in Fig.~\ref{fig2} (b). In CDW phase, we find that most odd-order HHG peaks can be observed, which is different from the fact that the peaks with number $3(2n+1)$ are suppressed in $24$A lattice. The results further support our proposal to use the odd- or even-order components to distinguish topological and local-order phases.

\begin{figure}[t]
\centering
\includegraphics[width=0.5\textwidth]{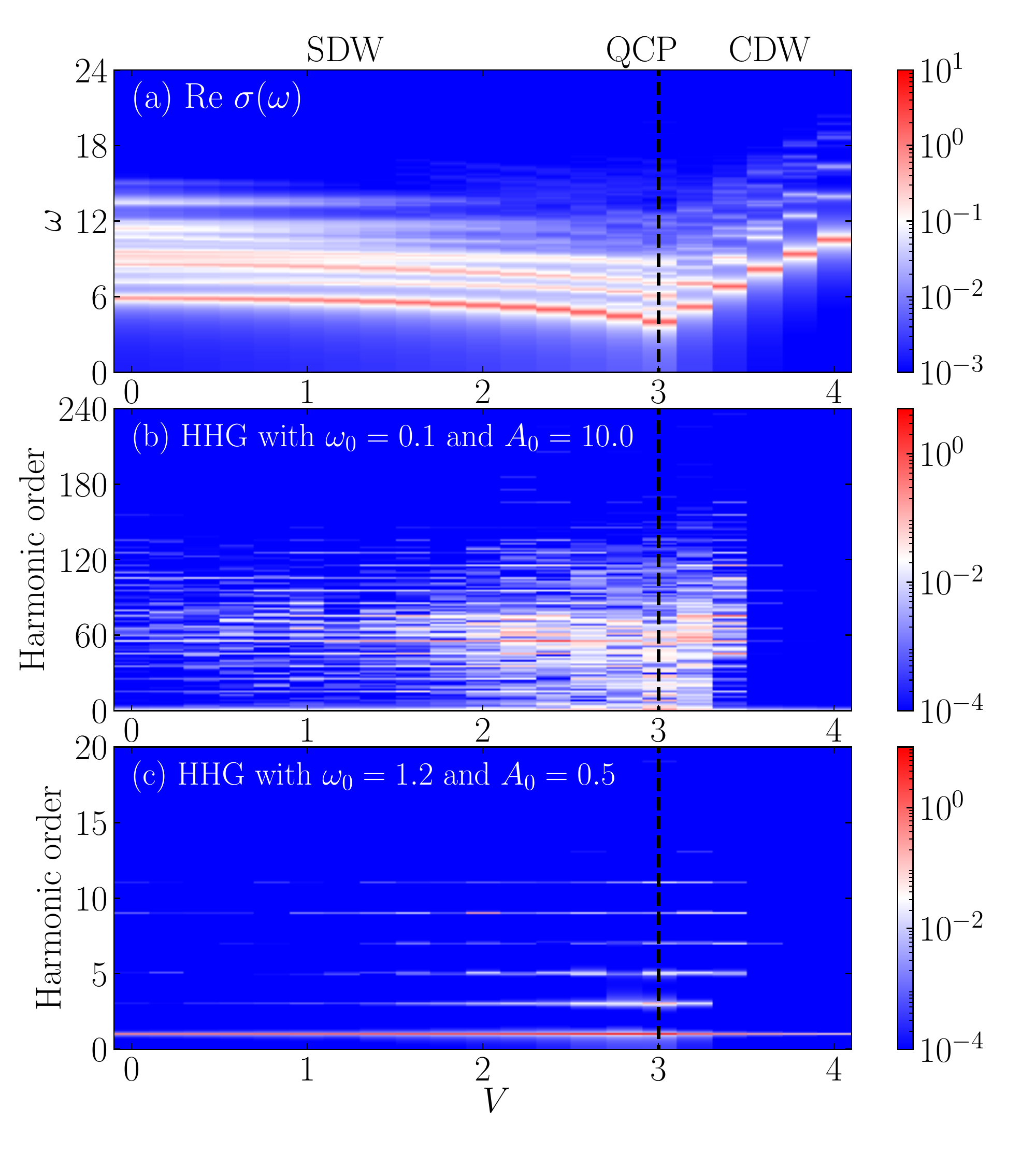}
\caption{
(a) Contour plots of the optical conductivity Re $\sigma(\omega)$ as a function of $\omega$ and the NN interactions $V$. Contour plots of HHG spectrum $|\langle j\rangle_{\omega}|^{2}$ as a function of $\omega/\omega_0$ and $V$, with $\omega_0=0.1$ and $A_0=10.0$ in (b) as well as $\omega_0=1.2$ and $A_0=0.5$ in (c). Other parameters of the two-leg extended Hubbard model and the external laser are set to be $U=9.0$ and $t_d=50.0$.
}
\label{fig3}
\end{figure}

%\begin{figure}[t]
%\centering
%\includegraphics[width=0.5\textwidth]{Figure_A5.pdf}
%\caption{
%The HHG spectrum $|\langle j\rangle_{\omega}|^{2}$ as a function of $\omega/\omega_0$ with $V=0.0$ (a), $V=5.0$ (b) and $V=6.0$ (c). \shao{Red lines emphasize the intensity difference of lower-order HHG ($3$rd and $5$th harmonic order).} Time profiles of $A(t)$ (red lines) and $\langle j\rangle_t$ (blue lines) with $V=0.0$ (d), $V=5.0$ (e) and $V=6.0$ (f). Other parameters of of the two-leg extended Hubbard model and the external laser are set to be $U=10.0$, $\omega_0=1.2$, $A_0=0.5$ and $t_d=50.0$.
%}
%\label{fig4}
%\end{figure}

\section{Supplemental Material (VI): HHG spectrum of the extended Hubbard model on a two-leg ladder.}

In this part, we choose a two-leg ladder with the lattice size $L = 2\times6=12$ to calculate the HHG spectrum of the extended Hubbard model. The periodic and open boundary conditions are applied along the leg and rung, respectively. The external laser to generate HHG is set to be linearly polarized along the leg and the optical conductivity Re $\sigma(\omega)$ is also defined along this direction. The on-site interaction $U=9$ and phase diagram bewteen the spin-density-wave (SDW) and charge-density-wave (CDW) states locates at $V\approx U/3=3.0$~\cite{Vojta1999}. We plot Re $\sigma(\omega)$ in equilibrium with changing the nearest-neighbour (NN) interaction $V$ in Fig.~\ref{fig3} (a), which shows a minium optical gap at $V=3.0$.

The parameters of the external laser are identical to those in 1D chain in the main text: $\omega_0=0.1$ and $A_0=10.0$ in Fig.~\ref{fig3} (b); $\omega_0=1.2$ and $A_0=0.5$ in Fig.~\ref{fig3} (c); $t_d=50.0$ for the both figures.
We can also observe an enhancement of the HHG intensity (especially for lower-order harmonics) when the system is approaching the critical point, attesting to the generality of our results.

\end{document}